\documentclass[12pt]{iopart}

\expandafter\let\csname equation*\endcsname\relax
\expandafter\let\csname endequation*\endcsname\relax

\usepackage{amsmath}
\usepackage{amssymb}
\usepackage{color}
\usepackage{verbatim}
\usepackage{array}
\usepackage{graphicx}
\usepackage{epsfig}
\usepackage{bm}
\usepackage[ragged]{sidecap}
\usepackage{dsfont}
\usepackage{ulem}
\usepackage{mathrsfs}

\newcommand{\ds}{\displaystyle}

\newcommand{\bra}[1]{\mathinner{\langle{#1}|}}
\newcommand{\ket}[1]{\mathinner{|{#1}\rangle}}
\newcommand{\braket}[2]{\langle #1|#2\rangle}
\newcommand{\comm}[2]{\left[#1,#2\right]}

\renewcommand{\vec}{\bm}
\newcommand{\ee}{\mathrm{e}}

\newcommand{\opvector}[1]{\hat{\mathbf{#1}}}
\newcommand{\opscalar}[1]{\hat{\textrm #1}}

\begin{document}

%=========Title=======================================================
\title[Quantum mechanical evolution operator]{Quantum mechanical evolution operator in the presence of a scalar linear potential: discussion on the evolved state}

%=========Authors=====================================================
\author{F Fratini$^{1,2,3}$ and L Safari$^{2}$}
\address{$^1$ Departamento de F\'isica, Universidade Federal de Minas Gerais, 30123-970 Belo Horizonte, Brazil}
\vspace{0.2cm}
\address{$^2$
Department of Physics, University of Oulu, Fin-90014 Oulu, Finland}
\address{$^3$
Institut N\'eel-CNRS, BP 166, 25 rue des Martyrs, 38042 Grenoble Cedex 9, France
}
\ead{
ffratini@fisica.ufmg.br , laleh.safari@oulu.fi
}

%=========Abstract====================================================
\begin{abstract}
We discuss the form of the wave-function of a state subjected to a scalar linear potential, paying special attention to quantum tunneling. We analyze the phases acquired by the evolved state and show that some of them have a pure quantum mechanical origin. In order to measure one of these phases, we propose a simple experimental scenario. We finally apply the evolution equations to re-analyze the Stern\&Gerlach experiment and to show how to manipulate spin by employing constant electric fields.
\end{abstract}

%=========Classification==============================================

\maketitle

%=========Text========================================================
\section{Introduction}

Scalar linear potentials are widely used in physics, as they can be generated by homogeneous irrotational fields, like electrostatic or gravitational fields, which are very typical in physical problems.
Moreover, linear potentials are of general use since they can approximate more sophisticated potentials for sufficiently small distances. Scalar linear potentials are also considered for quantum tunneling (e.g. the Sauter potential \cite{Cal1999}). Being able to rigorously evolve a quantum mechanical state subjected to a linear potential is therefore of fundamental and pedagogical importance.

The exact evolution of quantum systems subjected to linear and quadratic potentials has recently attracted some interest. 
The quantum propagator in the presence of a linear potential has been studied in several works \cite{Hol1997,Arr1996,Bro1994,Rob1996}.
The evolution operator related to the most generic time-dependent quadratic potential (with linear terms included) has been also analyzed in the literature by using quantum invariants \cite{Liu2004,Har2011}. 
Here, we re-derive the evolution operator in the presence of a scalar time-independent linear potential by using the Zassenhaus formula \cite{Que2004}. Our simple approach allows for several physical considerations that are laid out in the article.

In Sec. \ref{sec:Ev}, we show that the wave-function of a state evolved in the presence of a linear potential is given, up to a phase, by the free evolved wave-function (i.e., evolved with no potential) whose argument is shifted by a certain quantity which depends on the potential.
In Sec. \ref{sec:EvGauss}, we consider a gaussian wave-packet subjected to a linear potential. We pay some attention to the problem of quantum tunneling, or quantum diffusion, which is the phenomenon where a microscopic object (typically a particle or an atom) can penetrate a potential barrier whose height is larger than the object's kinetic energy \cite{Mohsen, Lau2000}. Since such phenomenon is forbidden by classical laws of mechanics, it is often referred to as a peculiar characteristic of quantum mechanics. 
We then move to discuss, in Sec. \ref{sec:PSG}, the form of the evolved state and the phases that it acquires. Some of these phases are shown to stem from the non-commutativity of momentum and position operators in quantum mechanics. A simple experimental scenario aimed at measuring one of those phases is proposed. 
In Sec. \ref{sec:SG}, as a pedagogic application of our evolution equations, we rigorously re-analyze the example of the Stern\&Gerlach (SG) experiment. 
In Sec. \ref{sec:SpinE}, we show how to manipulate spin of charged particles by using constant electric fields, instead of the more commonly used magnetic fields. 
Finally, a summary is given in Sec. \ref{sec:SumC}.

%----------------------------------------------------------------
\section{Evolution Operator and evolved wave-functions}
\label{sec:Ev}
We consider a potential of the form $V_0 x$, where $V_0$ is an arbitrary constant or any operator which commutes with momentum and position operators. Without restriction of generality and for simplicity, we consider only one dimension, which is the $x$ direction. The hamiltonian may be thus written as
\begin{equation}
\label{eq:Htot}
\opscalar H=\frac{\opscalar p^2}{2m}+V_0\opscalar x ~,
\end{equation}
where $\opscalar p$ is the linear momentum operator along the $x$ direction and $m$ is the particle mass. 
Since the hamiltonian is time-independent, we may straightforwardly write down the correspondent evolution operator from an initial time $t_i$ to $t$ \cite{Sak1994}:
\begin{equation}
\label{eq:UtotPrim}
\opscalar U(t,t_i)=e^{-\frac{i}{\hbar}\left(\frac{\opscalar p^2}{2m}+V_0\opscalar x\right)(t-t_i)} ~,
\end{equation}
where $\hbar$ denotes the reduced Planck constant. However, since the argument of the exponential in the equation above is made of non-commutative operators, its application to ket states is non-trivial. In order to rewrite \eqref{eq:UtotPrim} in a more manageable way, we make use of the Zassenhaus formula \cite{Que2004}:
\begin{equation}
\label{eq:Zass}
e^{\opscalar A+\opscalar B}=e^{\opscalar A}\,e^{\opscalar B}\,\prod_{i=2}^{\infty}e^{\opscalar C_i}~,
\end{equation}
where 
\begin{equation}
\begin{array}{lcl}
\opscalar C_2&=&\ds\frac{1}{2}\left[\opscalar B,\opscalar A\right]~,\\[0.4cm]
\opscalar C_3&=&\ds\frac{1}{3}\comm{\comm{\opscalar B}{\opscalar A}}{\opscalar B}\,+\,\frac{1}{6}\comm{\comm{\opscalar B}{\opscalar A}}{\opscalar A}~,\\[0.4cm]
\opscalar C_4&=&\ds\frac{1}{8}\left(
\comm{\comm{\comm{\opscalar B}{\opscalar A}}{\opscalar B}}{\opscalar B}
\,+\,
\comm{\comm{\comm{B}{A}}{A}}{B}
\right)\,+\,
\frac{1}{24}\comm{\comm{\comm{\opscalar B}{\opscalar A}}{\opscalar A}}{\opscalar A}
~,\\
...&&...~
\end{array}
\end{equation}
By choosing $\opscalar A\equiv-\frac{i}{\hbar}V_0\opscalar x\,(t-t_i)$, $\opscalar B\equiv-\frac{i}{\hbar}\frac{\opscalar p^2}{2m}\,(t-t_i)$ and by using $\comm{\opscalar p^2}{x}=\opscalar p\comm{\opscalar p}{x}+\comm{\opscalar p}{x}\opscalar p=-2i\hbar\opscalar p$, we readily get $\opscalar C_2=+\frac{i}{\hbar}\frac{V_0\opscalar p}{2m}(t-t_i)^2$, $\opscalar C_3=-\frac{i}{\hbar}\frac{V_0^2}{6m}(t-t_i)^3$, $\opscalar C_{i\ge4}=0$. Finally, by using the fact that $e^{\opscalar B}$ commutes with $e^{\opscalar C_2}$ and that $e^{\opscalar C_3}$ commutes with anything, we may write the (exact) evolution operator as
\begin{equation}
\label{eq:Utot}
\begin{array}{l}
\opscalar U(t,t_i)=
e^{-\frac{i}{\hbar}\frac{V_0^2}{6m}(t-t_i)^3}
e^{-\frac{i}{\hbar}V_0\opscalar x(t-t_i)}\, e^{+\frac{i}{\hbar}\frac{V_0\opscalar p}{2m}(t-t_i)^2}
\opscalar U_0(t,t_i)~,
\end{array}
\end{equation}
where
\begin{equation}
\label{eq:Ufree}
\opscalar U_0(t,t_i)=e^{-\frac{i}{\hbar}\frac{\opscalar p^2}{2m}(t-t_i)} 
\end{equation}
is the evolution operator in the free case, i.e. if there were no potential. Alternatively, by choosing $\opscalar B\equiv-\frac{i}{\hbar}V_0\opscalar x\,(t-t_i)$, $\opscalar A\equiv-\frac{i}{\hbar}\frac{\opscalar p^2}{2m}\,(t-t_i)$, one may analogously derive
\begin{equation}
\label{eq:Utot2}
\begin{array}{l}
\opscalar U(t,t_i)=\opscalar U_0(t,t_i)e^{+\frac{i}{\hbar}\frac{V_0^2}{3m}(t-t_i)^3}\, 
e^{-\frac{i}{\hbar}V_0\opscalar x(t-t_i)}e^{-\frac{i}{\hbar}\frac{V_0\opscalar p}{2m}(t-t_i)^2}~.
\end{array}
\end{equation}
Equation \eqref{eq:Utot} coincides with Eq. (2.6) in Ref. \cite{Arr1996}.

We notice that, for times much shorter than the characteristic time of interaction between potential and particle, 
we could neglect the terms $\sim(t-t_i)^2$ and $\sim (t-t_i)^3$ in the exponentials of Eqs. \eqref{eq:Utot2} and \eqref{eq:Utot}, in favor of the linear terms $\sim(t-t_i)$. 
The form thus obtained for the evolution operator $\opscalar U(t,t_i)$ would be equal to the one obtainable from Eq. \eqref{eq:UtotPrim} by considering as if the kinetic energy operator ($\frac{\opscalar p^2}{2m}$) and the potential energy operator ($V_0\opscalar x$) commuted. This means that, for short interaction times, the kinetic and the potential energy operators may be considered to approximately commute. This result is not unexpected, as it is indeed the basic step for the path integral formulation of quantum mechanics \cite{Alt2001}.

Next, we apply \eqref{eq:Utot} to an arbitrary initial state $\ket{\alpha(t_i)}$ defined at time $t_i$, so as to obtain the ket state at time $t$. By then multiplying by $\bra{x}$ from the left, we obtain the wavefunction of the state at time $t$ in the spatial representation:
\begin{equation}
\begin{array}{lcl}
\ds\Psi(x,t)&=& \braket{x}{\alpha(t)}=\ds\bra{x}\opscalar U(t,t_i)\ket{\alpha(t_i)}\\[0.4cm]
&=&\ds e^{-\frac{i}{\hbar}\frac{V_0^2}{6m}(t-t_i)^3}e^{-\frac{i}{\hbar}V_0 x(t-t_i)}\,
\bra{x}e^{+\frac{i}{\hbar}\frac{V_0\opscalar p}{2m}(t-t_i)^2}\ket{\alpha(t)}_0
\end{array}
\end{equation}
where we defined $\ket{\alpha(t)}_0=\opscalar U_0(t,t_i)\ket{\alpha(t_i)}$, which is just the state evolved by the free evolution operator \eqref{eq:Ufree}.
By using $\exp{\Big(-i\opscalar p \Delta x/\hbar\Big)}\ket{x}=\ket{x+\Delta x}$, which follows from the definition of momentum operator as generator of spatial translations \cite{Sak1994}, we finally obtain 
\begin{equation}
\label{eq:WFr}
\begin{array}{l}
\Psi(x,t)=\ds e^{-\frac{i}{\hbar}\frac{V_0^2}{6m}(t-t_i)^3} 
e^{-\frac{i}{\hbar}V_0 x(t-t_i)} \,\ds\Psi_0\left(x+\frac{V_0}{2m}(t-t_i)^2,t\right)~.
\end{array}
\end{equation}
It is important to notice that, in the equation above, the free evolved wave-function $\Psi_0$ can be of any form.
In the special case $\Psi_0$ is a plane-wave, then $\Psi$ will be a solution of the time-dependent Schr\"odinger equation related to the Hamiltonian \eqref{eq:Htot} \cite{Werner}. The modulus squared of the wave-function (i.e., the probability density) simply satisfies
\begin{equation}
\label{eq:SWFr}
\begin{array}{l}
\ds\left|\Psi(x,t)\right|^2= \left|\Psi_0\left(x+\frac{V_0}{2m}(t-t_i)^2,t\right)\right|^2~.
\end{array}
\end{equation}
We may obtain analogous relations in the momentum representation:
\begin{eqnarray}
\nonumber
\ds\tilde\Psi(p,t)&=& \braket{p}{\alpha(t)}=\ds\bra{p}\opscalar U(t,t_i)\ket{\alpha(t_i)}\\[0.4cm]
\label{eq:WFp}
&=&\ds e^{+\frac{i}{\hbar}\frac{V_0^2}{3m}(t-t_i)^3}e^{+\frac{i}{\hbar}\frac{V_0p}{2m}(t-t_i)^2}
\,\ds \tilde\Psi_0\Big(p+V_0(t-t_i),t\Big)~,\\[0.4cm]
\ds\left|\tilde\Psi(p,t)\right|^2&=&\ds \left|\tilde\Psi_0\Big(p+V_0(t-t_i),t\Big)\right|^2~,
\end{eqnarray}
where the relation $\exp{\Big(i\opscalar x \Delta p/\hbar\Big)}\ket{p}=\ket{p+\Delta p}$ has been used.

Equations \eqref{eq:WFr}, \eqref{eq:WFp} relate the wave-functions (in the spatial and linear momentum representations) of a general state evolved in the presence of a linear potential with the wave-functions of the same state evolved without the linear potential. 
The evolved wave-function is given, up to a phase, by the free evolved wave-function with its argument evolved following the classical equation of motion in the presence of the opposite potential.  For example, in the case of spatial representation, the argument is evolved following $x\to x+\frac{V_0}{2m}(t-t_i)^2$, while the classical evolution of the position given by the potential $V_0$ would be $x\to x-\frac{V_0}{2m}(t-t_i)^2$. Although this might seem counterintuitive at a first sight, in the next section we shall see that it is not. 

%-----------------------------------------------
\section{Evolution of a Gaussian wave-packet}
\label{sec:EvGauss}

Let us consider at initial time $t_i$ a Gaussian wave-packet with momentum mean value $p_0$, spatial mean value $x_0$ and standard deviation (or width) $\sigma$, which represents a realistic state in standard experiments:
\begin{equation}
\label{eq:Gauss}
\Psi^{G}(x,t_i)=\frac{1}{\pi^{1/4}\sigma^{1/2}}e^{\frac{i}{\hbar}p_0(x-x_0)-\frac{(x-x_0)^2}{2\sigma^2}}\,\equiv\,
\braket{x}{G}~.
\end{equation}
We shall denote with $\Psi^{G}_0(x,t)$ the free evolved Gaussian wave-packet at time $t$, i.e. $\Psi^{G}_0(x,t)=\bra{x}\opscalar U_0(t,t_i)\ket{G}$. 
Using Eq. \eqref{eq:SWFr} with the Gaussian state \eqref{eq:Gauss} (i.e., replacing $\Psi_0(x,t)\to\Psi^{G}_0(x,t)$), the modulus squared of the wave-packet at time $t$ is of the form:
\begin{equation}
\label{eq:Gstate}
\begin{array}{lcl}
\Big|\Psi^G(x, t)\Big|^2&=& 
\Big| \Psi_0^G(x+\frac{V_0}{2m}(t-t_i)^2, t) \Big|^2\propto
e^{-\frac{\left(x+V_0\Delta t^2/2m-(x_0+p_0\Delta t/m)\right)^2}{\left|\sigma(\Delta t)\right|^2}}\\[0.4cm]
&\propto& e^{-\frac{\left(x-(x_0+p_0\Delta t/m-V_0\Delta t^2/2m)\right)^2}{\left|\sigma(\Delta t)\right|^2}}
\end{array}
\end{equation}
where $\Delta t=t-t_i$, and $\sigma(\Delta t)$ denotes the spatial width of the Gaussian wave-packet free evolved for a time $\Delta t$. The term $p_0\Delta t/m$ comes from the free evolution of the Gaussian wave-packet, as it could be expected. The detailed expression of $\sigma(\Delta t)$ and $\Psi^{G}_0(x,t)$ can be found in standard textbooks \cite{Merz}. Evidently, equation \eqref{eq:Gstate} describes the probability density of a Gaussian wave-packet whose spatial mean value obeys non-relativistic classical kinematics. In other words, The spatial mean value has been subjected to a constant acceleration equal to $-\frac{V_0}{m}$ for a time interval $\Delta t$. This result is also in line with Ehrenfest's theorem for the mean values \cite{Sak1994}. In conclusion, the fact that the wave-packet argument follows the classical equation of motion in the presence of the opposite potential ensures that the wave-packet mean value follows the classical equation of motion with the correct potential. The same analysis can be conducted in the momentum representation, with the same classical results. We thus conclude that the evolution equations \eqref{eq:WFr} and \eqref{eq:WFp} are physically plausible, when analyzed from the point of view of classical kinematics. We furthermore notice that the presence of the linear potential does not affect the spatial width $\sigma$, which is rather fully determined by the free evolution in time. 

\begin{figure}
\centering
\includegraphics[scale=1.2]{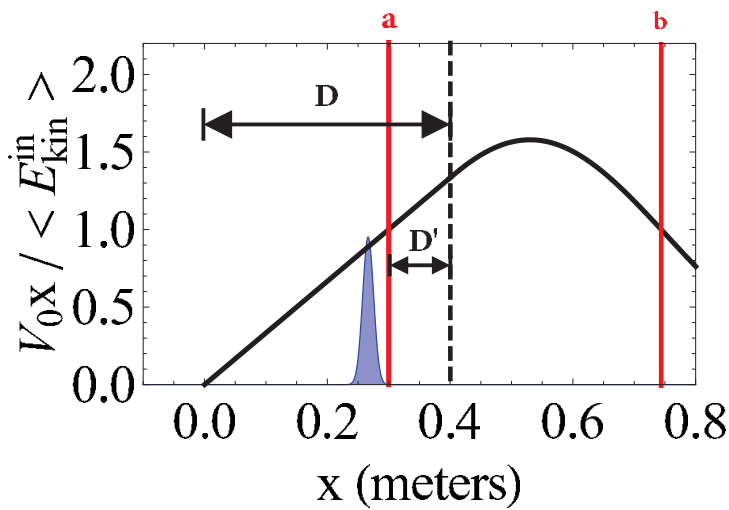}
\caption{A gaussian wave-packet (blue curve) hits a potential barrier (black curve) whose initial part (D) can be approximated to linear. The potential barrier is in units of the initial average value for the kinetic energy of the wave-packet ($<E_{kin}^{in}>=p_0^2/(2m)$). The red solid vertical bars denote the classical turning points (a, b). The length D$'$ is chosen to be much larger than half wave-packet spatial width as given at the time $t_a$ (D$'$$\,\gg\frac{\sigma(t_a-t_i)}{2}$). Here $t_a=p_0/V_0$ is the time when the wave-packet mean value is at the turning point $a$.
For these settings, as a consequence of the evolution given by Eq. \eqref{eq:Gstate}, the wave-packet will be wholly pushed backwards and no tunneling will be permitted.
}
\label{fig:fig1}
\end{figure}

Equation \eqref{eq:Gstate} may be also read in the following way: When the whole wave-packet is subjected to a linear potential, the (modulus squared of the) wave-packet follows the classical kinematics evolution. Now, let us consider a gaussian wave-packet that hits a potential barrier whose initial part can be approximated to linear. This initial (approximately) linear part of the potential shall be called D. The maximum value of the potential in D is supposed to be higher than the initial average kinetic energy of the wave-packet, where this latter is $<E_{kin}^{in}>=p_0^2/(2m)$. For convenience, this situation is depicted in Fig. \ref{fig:fig1} and also showed in the animation available on line as supplementary material (in the animation, also the evolution of the wave-packet in the free case is displayed for comparison). Let us further suppose that the difference between the furthermost point of D and the first classical turning point (a) is much bigger than half wave-packet spatial width as given at the time $t_a$ (D$'$$\,\gg\frac{\sigma(t_a-t_i)}{2}$). Here $t_a=p_0/V_0$ is the time when the wave-packet mean value is at the turning point $a$. In other words, we suppose that the DeBroglie wave-length of the wave-packet is small compared to the characteristic distance over which the first derivative of the potential varies appreciably \footnote{This is a less stringent approximation than the assumption for the validity of WKB approach. Within WKB approach, which is widely used in quantum tunneling problems, the DeBroglie wave-length of the packet is considered to be small compared to the characteristic distance over which the potential varies appreciably \cite{Sak1994}.}. In these chosen settings, the whole wave-packet will be subjected to the same linear potential up to the turning point a. The evolution given by Eq. \eqref{eq:Gstate} then dictates that the wave-packet will be wholly pushed backwards, as one would expect from the classical point of view. Consequently, {\it no tunneling will be permitted to the wave-packet}. Therefore, in order to have any chance for quantum tunneling, the wave-packet spatial width must be somewhat larger or comparable to D$'$ ($\sigma\gtrsim$ D$'$), so that the evolution given by \eqref{eq:Gstate} may not be applicable. 

In the animation available on line as supplementary material, we chose the mean position, the standard deviation and the velocity to be $0$ m, $0.2$ cm and $1$ m/sec, respectively, at time $t=0$. The strength of the potential is chosen to allow the wave-packet to travel for a length of $0.3$ meters before reaching the classical turning point. At the classical turning point, the spatial width will have grown of only 10\% (see animation). For the best comparison, we adopted the same length unit of the animation for the abscissa in Fig. \ref{fig:fig1}. Other settings for smaller length scale could be analogously applied.

Based on the above considerations, we argue that quantum tunneling reflection and transmission coefficients should directly depend on the spatial width of the wave-packet. 
However, although studies on tunneling with Gaussian wave-packets have been made in literature (see Ref. \cite{Stamp1996} and references therein), to the best of our knowledge no direct relation between tunneling coefficients and spatial width of the wave-packet has been suggested. 
In fact, transmission coefficients in quantum tunneling are not normally given as dependent on the wave-packet spatial width but rather as solely dependent on the energy of the particle (E) and the thickness of the barrier. For instance, within WKB approximation, which is the most widely used approach for solving tunneling problems, the transmission coefficient is given by $\ds T(E)=e^{-2\sigma_R}/\left(1+e^{-2\sigma_R}/4\right)^2$, where $\sigma_R=\int_a^b\sqrt{\frac{2m}{\hbar^2}\left(V(x)-E\right)}\,dx \,>\,0$, and $a$, $b$ are the classical turning points (i.e., $L=b-a$ is the classically forbidden region) \cite{Mohsen}. 
An experimental assessment of the dependence of tunneling coefficients on the spatial width would thus be desirable. 

The direct dependence of tunneling coefficients on the spatial width of the wave-packet could have application in many areas of science: Quantum tunneling could be enhanced or suppressed by controlling the spatial width of the state, instead of controlling the energy of the state or the environment surrounding it \cite{Gri1998, Cal1981}.

Unfortunately, we cannot find here an explicit expression for transmission and reflection coefficients for a realistic potential barrier with the present quantum mechanical formalism. This is because the Zassenhaus formula does not converge for potentials of order higher than linear, and because a realistic potential barrier cannot be represented by a linear function. Nonetheless, any potential barrier can be approximated to linear for short distances. Based on this, our claim that the tunneling coefficients should depend on the spatial width of the wave-packet holds. In view of the fact that the wave-packet spatial mean value ($x_0$) follows the classical equation of motion, our conjecture is that the fraction of the wave-packet beyond the potential barrier at the classical turning point plays leading role in determining transmission coefficient in quantum tunneling: Given wave-packets with the same linear momentum mean value, those wave-packets with larger spatial width will tunnel more efficiently. This can be simply checked by preparing Gaussian wave-packets and delaying the arrival of some of them to the potential barrier. The spatial width $\sigma(\Delta t)$ of the wave-packets increases during the free evolution. Thus the delayed wave-packets will have larger spatial width with respect to the non-delayed wave-packets.

%-----------------------------------------
\section{Discussion on the phases: Phase Shift Generator}
\label{sec:PSG}
We here discuss more extensively the expression for the evolution operator in Eq. \eqref{eq:Utot}. 

The first $\left(-\frac{i}{\hbar}\frac{V_0^2}{6m}(t-t_i)^3\right)$ and third $\left(+\frac{i}{\hbar}\frac{V_0\opscalar p}{2m}(t-t_i)^2\right)$ phases which multiply the free evolution operator from the left in \eqref{eq:Utot} stem directly from the non-commutativity between momentum and position operators, and are thus purely quantum mechanical corrections. 
On the other hand, the second phase $\left(-\frac{i}{\hbar}V_0\opscalar x(t-t_i)\right)$ and the free evolution operator itself ($\opscalar U_0(t,t_i)$) somehow represent the evolution given by the potential and kinetic energy gained by the traveling particle, respectively.
If position and momentum operators commuted, then only these two latter terms would be present.
%Since this is not the case, even if only these two parts were considered, the evolution given by either operators would not simply result in an overall phase.
 
It is somewhat interesting that the phase $\frac{V_0\opscalar p}{2m\hbar}(t-t_i)^2$ is directly responsible in equation \eqref{eq:WFr} for the shift in the argument of the free evolved wave-function in the spatial representation, such shift being $V_0/(2m)\,(t-t_i)^2$. In fact, that shift, together with the shift $p_0\Delta t/m$ given by the free evolution, gives rise to the classical motion of the spatial mean value of the state (see previous section). Therefore, the non-commutativity of momentum and position operators turns out to be effectively responsible for the non-relativistic classical motion of the spatial mean value of the wave-packet: $x_0\to x_0+p_0\Delta t/m-V_0\Delta t^2/2m$. We may furthermore notice that the corresponding shift in the momentum representation, which is responsible for the classical motion of the linear momentum mean value, is directly given by the phase containing the linear potential, $\left(-\frac{i}{\hbar}V_0\opscalar x(t-t_i)\right)$. We may therefore consider the term $\propto\frac{V_0\opscalar p}{2m}$ as a linear potential in the momentum space. In other words, the term $\propto\frac{V_0\opscalar p}{2m}$ may be considered the dual of the potential $V_0\opscalar x(t-t_i)$. The former is generated by the presence of the latter because the latter does not commute with the free Hamiltonian. The presence of both potentials gives symmetry to the evolution of the state in spatial and momentum representations and ensures that in both representations the mean value is evolved following the non-relativistic classical motion.

On the other hand, the phase $-\frac{1}{\hbar}\frac{V_0^2}{6m}(t-t_i)^3$ does not play any role on determining the classical evolution of the state. Such a phase originates from the second (and last) expansion term of the Zassenhaus formula, and it is therefore a higher correction with respect to other terms. Indeed, this phase would be the only one missing if we replaced, in the free plane wave $e^{\frac{i}{\hbar}\left(px-\frac{p^2}{2m}(t-t_i)\right)}$, the classical transformations $x\to x-\frac{V_0}{2m}(t-t_i)^2$, $p\to p-\frac{V_0}{m}(t-t_i)$, as one would do as a first attempt to guess the wave-function of a state subjected to a linear potential.
An experiment aimed at ascertaining the existence of this last phase would thus probably be a useful test for quantum mechanics. To this aim, here we sketch a simple experimental scenario which permits such a measurement.

\begin{figure}
\centering
\includegraphics[scale=0.5]{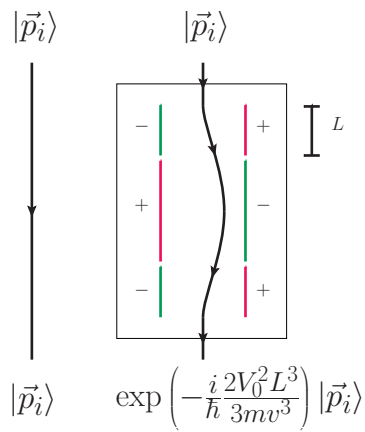}
\caption{Sketch of a simple experimental scenario which allows to generate a phase shift using the evolution operator in Eqs. \eqref{eq:Utot}, \eqref{eq:Utot2}.
}
\label{fig:fig2}
\end{figure}

Let us consider two electron beams along the $z$ direction, where the electrons are in phase one with another \cite{Ton2985}. One of the two beams is accelerated and subsequently decelerated along an axis orthogonal to the beam direction, for instance $x$. In order to apply Eqs. \eqref{eq:Utot}-\eqref{eq:Utot2}, the acceleration and deceleration must be due to a linear potential. As showed in Fig. \ref{fig:fig2}, this could be realized, for instance, by a series of three capacitors of lengths $L=v\Delta t$, $2L$ and $L$, where $\Delta t$ is the time the electron spends in the first capacitor and $v$ is the beam velocity along $z$ ($L$ must here be much larger than the spatial width of the electron state). For this example, we must consider the three dimensional generalization of Eq. \eqref{eq:Utot}, where $\opscalar p^2$ is replaced by $\opvector p^2=\opscalar p_x^2+\opscalar p_y^2+\opscalar p_z^2$. Since position and momentum operators along different directions commute, such a replacement can be safely made.
By applying such generalized evolution operator to the initial electron state $\ket{\vec p_i}\simeq\ket{p_x=0, p_y=0, p_z=p_0}$, the electron state after the electrostatic deflection (which lasts for a time $4\Delta t$) can be easily calculated to be $e^{-\frac{i}{\hbar}\frac{2V_0^2 L^3}{3m\,v^3}}\opscalar U_0(4\Delta t)\ket{\vec p_i}\equiv e^{-\frac{i}{\hbar}\frac{2V_0^2 L^3}{3m\,v^3}} \ket{\vec p_i}$. We have here redefined $\opscalar U_0(4\Delta t)\ket{\vec p_i}\equiv \ket{\vec p_i}$ since the phase given by the free evolution operator is shared by both beams and therefore not measurable. Thus, upon passing the capacitors, the beam acquires a phase-shift equal to $-\frac{1}{\hbar}\frac{2V_0^2L^3}{3m v^3}$ with respect to the other (non-deflected) beam. Such phase shift can be measured when the beams are recombined. In what follows, we shall denote the experimental apparatus sketched in Fig. \ref{fig:fig2} as Phase Shift Generator (PSG).

From the above considerations we see that, when different beams are subjected to different accelerations, a phase difference proportional to the potential-difference squared may appear, if the potentials responsible for the accelerations can be approximated to linear. Therefore, Eqs. (\ref{eq:Utot}), (\ref{eq:Utot2}) and Fig. \ref{fig:fig2} might be also useful for estimating the loss of coherence in dealing with charged particles.

Phase shifts of quantum states have been very useful in physics and are object of current research and debate (e.g., the gravitational phase shift \cite{COW,Lit} and the Gouy phase \cite{G1,G2}). Along the same lines, ascertaining the existence of the phase shift generated by the PSG would be interesting for testing quantum mechanics and might also have several applications. We shall see in Sec. \ref{sec:SpinE} how such phase could be for example used in quantum information and spintronics.

%----------------------------------------------
\section{Application to Stern\&Gerlach experiment}
\label{sec:SG}

The SG experiment \cite{SG1922} is rightfully considered of fundamental and pedagogical importance for understanding quantum mechanics. The particles injected in the SG apparatus are subjected to a linear potential. The SG apparatus is therefore probably the best example for a clear application of equations \eqref{eq:Utot}-\eqref{eq:WFp}. The following brief analysis is also motivated by the fact that in textbooks the SG apparatus is normally explained with intuitive, semi-classical arguments (e.g., see in Ref. \cite{Sak1994}), while in literature it is more rigorously explained with an involved quantum mechanical formalism \cite{Scu1987}.

In the SG experiment, Silver atoms are injected in the apparatus \cite{SG1922}. Out of the 47 electrons of the Silver atom, only the outermost electron contributes to the atomic spin, if we neglect the nuclear contribution (which is irrelevant to our discussion). Therefore it is common to consider the spin state of such electron as characterizing the spin state of the whole atomic system. Since no atomic excitations are to be considered, we may disregard any atomic internal structure.  The potential for a SG whose magnetic field is along $x$ is $\opscalar V_{SG}\simeq-\left(\frac{\ee\hbar}{2m_e}B_0\right)\opscalar x\,\hat \sigma_x$, where $\ee$ and $m_e$ are the electric charge and mass respectively, $\hat \sigma_x$ is the Pauli spin operator along the $x$ direction, while $B_0$ is the strength of the magnetic field. The state of the atoms before entering the SG is completely mixed. Thus, it may not be described by a ket state, but rather it can be described by the following density operator \cite{Sak1994, Fratini2011}:
\begin{equation}
\begin{array}{lcl}
\hat \rho(t_i)&=&\ds\ket{\vec p}\bra{\vec p}\,\otimes\,
\frac{1}{2}\Big(\ket{S_x,+}\bra{S_x,+} + \ket{S_x,-}\bra{S_x,-}\Big)~,
\end{array}
\end{equation}
where $\vec p=(0,0,p_0)$ and $\ket{S_x, \pm}$ are spin-1/2 states along the $x$ direction. Setting $V_0\to \opscalar V_0 \equiv -\left(\frac{\ee\hbar}{2m_e}B_0\right)\,\hat \sigma_x$ in Eq. \eqref{eq:Utot} (which is allowed, since $\hat \sigma_x$ commutes with any of the operators $\opscalar p_x$, $\opscalar p_y$, $\opscalar p_z$, $\opscalar x$, $\opscalar y$, $\opscalar z$), the evolution of the density operator can be easily computed \cite{Balashov}:
\begin{equation}
\begin{array}{lcl}
\hat \rho(t)&=&\opscalar U(t, t_i)\hat \rho(t_i)\opscalar U^\dag(t, t_i)\\[0.4cm]
&=&\ds\frac{1}{2}\Big(
\ket{\vec p_+}\bra{\vec p_+}\,\otimes\,\ket{S_x,+}\bra{S_x,+}  \;+\;\ket{\vec p_-}\bra{\vec p_-}\,\otimes\,\ket{S_x,-}\bra{S_x,-}\Big)~,
\end{array}
\end{equation}
where $\vec p_+=\big(+\Delta p, 0, p_0\big)$ and $\vec p_-=\big(-\Delta p, 0, p_0\big)$, while $\Delta p=\frac{\ee\hbar}{2m_e}B_0(t-t_i)$. The probability density of measuring the generic state $\ket{\vec p'}\bra{\vec p'}$ after an interaction time $(t-t_i)$ is therefore
\begin{equation}
\Tr\Big[\ket{\vec p'}\bra{\vec p'}\hat \rho(t)\Big]=\frac{1}{2}\,\delta(\vec p'-\vec p_\pm)~,
\end{equation}
which is the well-known SG outcome.

%--------------------------------------------------------
\section{Manipulating spin by employing electric fields}
\label{sec:SpinE}

Manipulating spin by means of electric fields is currently subject of applied research \cite{Hon2013,Tsc2006,Jan2010,Rov2010,Han2008}. The advantage of controlling spin by using electric rather than magnetic fields is that the former are easier to generate. Moreover, they allow for controlling spins independently one from another, which is a requirement for building quantum computers \cite{Duc2006,Now2007}.
Our goal here is to show how to manipulate spin of charged particles by employing constant electric fields. This will be achieved by combining a spin-beam splitter with a PSG apparatus. For our purposes and for simplicity, we will consider electrons and we will use the SG apparatus as spin-beam splitter. A brief discussion on the feasibility and on possible extensions is laid out at the end of the present section.

\begin{figure}[!t]
\centering
\includegraphics[scale=0.5]{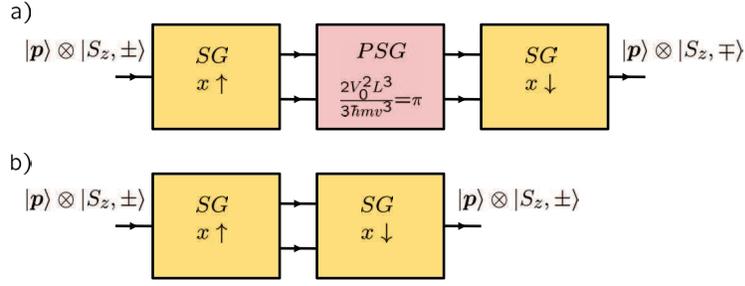}
\caption{a) A spin flipper built by employing SG and PSG apparatuses.
b) By removing the PSG, the spin is not flipped. 
}
\label{fig:fig3}
\end{figure}

We prepare an electron with, for example, momentum and spin along $z$. Its state is therefore described by $\ket{\vec p}\otimes\ket{S_z,\pm}$, with $\vec p=(0, 0, p_0)$. We let such electron sequentially pass through a SG, then through a PSG, and finally through a second SG. While the first SG is set along $x$ direction, the second SG is set along $-x$ direction. This gedanken experiment is sketched in Fig. \ref{fig:fig3} (panel a). During the several steps, by using Eq. \eqref{eq:Utot} we find that the electron state is given by (up to an overall phase):
\begin{equation}
\begin{array}{l}
\ket{\vec p}\otimes\ket{S_z,\pm}=\ds\frac{1}{\sqrt{2}}\ket{\vec p}\otimes\Big( 
\ket{S_x,+}
\pm
\ket{S_x,-}
\Big)\\[0.4cm]
\;\xrightarrow{SG\uparrow}\ds
\frac{1}{\sqrt{2}}\Big( 
\ket{\vec p_+}\otimes\ket{S_x,+}
\pm
\ket{\vec p_-}\otimes\ket{S_x,-}
\Big)\\[0.4cm]
\;\xrightarrow{PSG}\ds
\frac{1}{\sqrt{2}}\Big( 
\ket{\vec p_+}\otimes\ket{S_x,+}
\pm
e^{-\frac{i}{\hbar}\frac{2V_0^2L^3}{3m v^3}}\ket{\vec p_-}\otimes\ket{S_x,-}
\Big)\\[0.4cm]
\;\xrightarrow{SG\downarrow}\ds
\frac{1}{\sqrt{2}}\ket{\vec p}\otimes\Big( 
\ket{S_x,+}
\pm
e^{-\frac{i}{\hbar}\frac{2V_0^2L^3}{3m v^3}}\ket{S_x,-}
\Big)~,
\end{array}
\end{equation}
where $\vec p_{\pm}$ has been defined previously and contains the linear momentum shift given by the Lorentz force exerted on the electron in the intermediate steps.
By setting appropriate values for the PSG parameters, the wished final electron spin state is obtained. In particular, by setting $\frac{2V_0^2L^3}{3\hbar m v^3}=\pi$, the initial spin state is reversed.
By removing the PSG, the spin is not flipped (Fig. \ref{fig:fig3}, panel b). This entails that it is the electric field in PSG that is responsible for the spin flip, while the magnetic field in the SG apparatuses is just used to feed the PSG. 

The employment of a SG apparatus to split electron beams of different spins has been widely discussed in the past. Since the beginning ot the last century, it has been several times argued that the SG apparatus cannot successfully split electron beams. Conversely, more recently this viewpoint has been confuted and an effective spin-beam splitter for electrons using SG has been proposed (see Ref. \cite{Bate1997} and references therein). Here, we used the SG apparatus for simplicity, but any spin-beam splitter would work and would produce the shown results. Spin-beam splitters of nano- and meso-scopic dimensions have been in fact realized for electrons (e.g., \cite{Ch2008,Pe2006,X2006,Fre2003}). Our scenario for spin manipulation is therefore feasible with the current state-of-the-art technology.
%Alternatively, one could prepare single-ionized He atoms instead of electrons to use with the SG spin-beam splitter. The charge of such atoms is one electronic charge. On the other hand, the atomic spin is 1/2 and is wholly given by the electronic spin, since the nuclear spin is zero.

A microscopic realization of PSG jointly with a spin-beam splitter may be applied, for instance, in quantum information and spintronics (including atomtronics, if charged atoms are employed) \cite{Wol2001,Zu2004,Pep2009,Ru2004}. In fact, as showed in this section, such a combination works as a gate $\ket{0}+\ket{1}\to\ket{0}+e^{i\mu}\ket{1}$, where $\mu$ is any wished phase.

%-----------------------------------
\section{Summary}
\label{sec:SumC}

In summary, we used the Zassenhaus formula to re-derive the quantum mechanical evolution operator in the presence of a scalar linear potential. We discussed the form of the wave-function of the evolved state paying special attention to quantum tunneling. We then analyzed the phases that the evolved state acquires. We proposed an experimental scenario for measuring one particular phase given by the non-commutativity of momentum and position operators. We applied the evolution equations to rigorously re-analyze the Stern\&Gerlach experiment and to show how to manipulate spin by using constant electric fields.

%The possibility to manipulate spin by using electric fields instead of magnetic fields, as showed in Sec. \ref{sec:SpinE}, might also have some relevance in medical and material diagnostics \cite{med}. 

\ack
F.F. acknowledges Funda\c{c}\~ao de Amparo \`a Pesquisa do estado de Minas Gerais (FAPEMIG) and Conselho Nacional de Desenvolvimento Cient\'ifico e Tecnol\'ogico (CNPq).
L.S. and F.F. acknowledge support by the Research Council for Natural Sciences and Engineering of the Academy of Finland.
F.F. is thankful to Prof. S. A. Werner for suggestions and for kindly providing the slides of his talk given at NIST (op. cit.).
F.F. is thankful to Hakob Avetisyan for useful discussions.
The authors are thankful to an anonymous referee for his/her valuable comments and for having suggested important references to include.

\section*{References}

%=========Bibliography========================================================

\end{document}